\documentclass[aps,pra,superscriptaddress,twocolumn]{revtex4}
\usepackage{amsmath}
\usepackage{amsfonts}
\usepackage{graphicx}
\usepackage{psfrag}

\begin{document}

\title{Scattering of charged particles on two spatially separated time-periodic optical fields}
\author{L\'{o}r\'{a}nt Zs.~Szab\'{o}}
\affiliation{Department of Theoretical Physics, University of Szeged, Tisza Lajos k\"{o}r%
\'{u}t 84, H-6720 Szeged, Hungary}
\author{Mih\'{a}ly G.~Benedict}
\affiliation{Department of Theoretical Physics, University of Szeged, Tisza Lajos k\"{o}r%
\'{u}t 84, H-6720 Szeged, Hungary}
\author{P\'{e}ter F\"{o}ldi}
\affiliation{Department of Theoretical Physics, University of Szeged, Tisza Lajos k\"{o}r%
\'{u}t 84, H-6720 Szeged, Hungary}
\affiliation{ELI-ALPS, ELI-HU Non-profit Ltd., Dugonics t\'{e}r 13, H-6720 Szeged, Hungary}

\begin{abstract}
We consider a monoenergetic beam of moving charged particles interacting with two separated oscillating electric fields. Time-periodic linear potential is assumed to model the light-particle interaction using a non-relativistic, quantum mechanical description based on Gordon-Volkov states. Applying Floquet theory, we calculate transmission probabilities as a function of the laser field parameters. The transmission resonances in this Ramsey-like setup are interpreted as if they originated from a corresponding static double potential barrier with heights equal to the ponderomotive potential resulting from the oscillating field. Due to the opening of new "Floquet channels," the resonances are repeated at input energies when the corresponding frequency is shifted by an integer multiple of the exciting frequency. These narrow resonances can be used as precise energy filters. The fine structure of the transmission spectra is determined by the phase difference between the two oscillating light fields, allowing for the optical control of the transmission.
\end{abstract}

\maketitle

\section{Introduction}

Optical control of quantum mechanical particles offers a wide variety of promising applications, including ultrafast electronics \cite{krausz2009,Krausz2014,Gruber2016}, imaging \cite{feist2017,zewanat2009} or quantum computation \cite{Kim2011,Liu2010}. Among the first phenomena of describing the coupling of free electrons to light was the Kapitza-Dirac effect in the 1930s \cite{KD1933}. In this elastic process, diffraction of electrons is observed in a standing light wave which acts as an effective diffraction grating \cite{GRP_KD86,FAB_Nat_KD01}. Beams of electrons can also be manipulated by optical fields \cite{deAbajo2010,Echternkamp2016,Feist2015,Kozak2017,breuer2013}, while the properties of oscillating plasmonic near-fields can be probed by measuring electron spectra from nanostructures \cite{Dombi2013,Racz2017}. Recently, photon-induced near-field electron microscopy revealed that the initial kinetic energy distribution of short electron pulses broadens through induced photon sidebands \cite{pinemtheoexp,zewanat2009}.

For strong excitations, the highly inelastic photon-induced processes that involve the absorption/emission of one or a few photons can be appropriately described by using classical, periodic fields. In this intensity regime, Floquet theory is proved to be one of the most efficient methods. Although in this case the exciting field that oscillates with a frequency of $\omega$ is not quantized, the corresponding quasienergies are separated by integer multiples of $\hbar\omega.$ This means that for an inelastic process, the material response will contain frequency components that are integer multiples of $\omega,$ and, e.g., in transport processes, the transmitted energy spectrum will contain sidebands around the input energy. These sidebands correspond to Floquet channels with energy shifts $n\hbar\omega,$ where $n$ is integer.

Interferometry induced by spatially separated electromagnetic fields is a very important tool for the control of quantum mechanical particles, see e.g.~\cite{H13}. Here, we present a theoretical description of a Ramsey-like setup, where, instead of being bound to nuclei, {\em free} charged particles interact with two separated periodic electric fields in a non-relativistic framework. In more detail, similarly to Ref.~\cite{Echternkamp2016}, we describe the scattering of a monoenergetic particle wave on two localized optical fields. However, the main focus is on the calculated transmission spectra which are thoroughly investigated. Resonant tunneling is observed which is also known in static scattering problems, see e.g., \cite{kaminski2012,aktas2016,Anwar1989} in the context of nanostructures. Let us recall that in the vicinity of metallic nanoparticles the net electromagnetic field can become localized (e.g., in Ref.~\cite{kruger2011} a diameter of $\sim10\,nm$ was reported for the case of a nanoscale tip) and the space dependence of the field can be neglected. Motivated by this, we use dipole approximation, i.e, assume that the field has only time dependence.

As a first approach, we create a static scattering model, where we consider a rectangular potential barrier with height of the ponderomotive potential $U_{p}$. Its transmission spectrum is a good approximation for the time-dependent problem. We also investigate the induced photon sidebands and the space and time dependent probability current density.

Our paper is organized as follows: in Secs.~II and III we describe the theoretical framework with Gordon-Volkov states and derive the wave equations using Floquet theory. In Sec.~IV we show that a static scattering model can be thought as a good approximation for the time-dependent model regarding the transmission spectra. Transmission resonances are analyzed through induced photon sidebands and through space and time dependent probability current for various parameter ranges. We close our paper by summarizing our findings and drawing conclusions in Sec.~V.

\section{Model}

%%%%%%%%%%%%%%%%%%%%%%%%%%%%%%%%%%%%%%%%%%%%%%%%%%%%%%%%%%%%%%%%%%%%%%%%%%%%%
\begin{figure}[tbh]
\includegraphics[width=8.0cm]{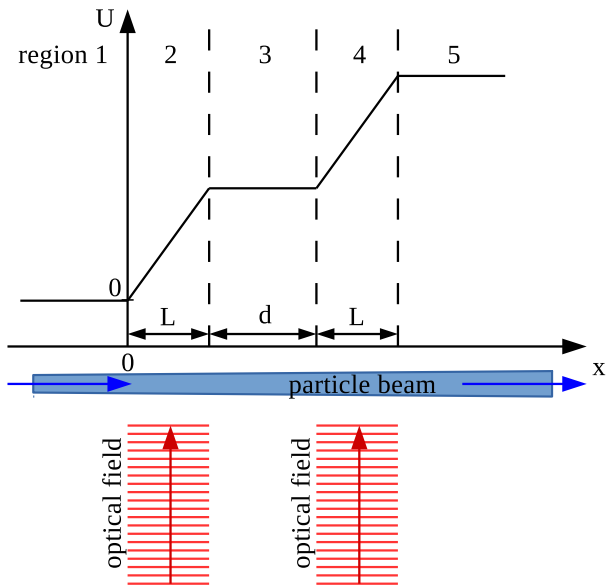}
\caption{Schematic view of the setup we consider. In region 1, we have an incident monoenergetic free particle wave propagating towards the oscillating time-dependent potential localized in a finite region with length $L$, inducing reflected and transmitted waves in region 1 and 5, respectively. In the interaction regions (2 and 4), superpositions of
Gordon-Volkov states [see Eq.~(\protect\ref{volk})] are present.}
\label{fig1}
\end{figure}
%%%%%%%%%%%%%%%%%%%%%%%%%%%%%%%%%%%%%%%%%%%%%%%%%%%%%%%%%%%%%%%%%%%%%%%%%%%%%%
We consider a one dimensional non-relativistic scattering model, where a beam of monoenergetic free charged particles is assumed to interact with two spatially separated linearly polarized time-periodic electric fields, see Fig.~\ref{fig1}. The direction of the matter wave propagation is chosen to be the $x$ axis, which, for the sake of simplicity, is divided into five regions. In region 1, the potential is zero, the Hamiltonian reads
\begin{equation}
H_{1}=\frac{p^{2}}{2m},
\label{H1}
\end{equation}
where $p=-i\hbar \,\partial /\partial x$ is the canonical momentum operator. We shall assume here an incident plane wave with energy $E_{0}$ and the corresponding momentum $\hbar k_{0}=\sqrt{2mE_{0}}$:
\begin{equation}
\psi_{in}(x,t)=e^{ik_{0}x-\frac{i}{\hbar}E_{0}t},
\label{psiin}
\end{equation}
which is a particular solution of the time-dependent Schr\"{o}dinger equation generated by $H_{1}$. The charged particles are assumed to interact with the laser field in regions 2 and 4 inducing free reflected and transmitted waves in regions 1 and 5, respectively, and both left and right propagating waves in region 3. In more detail, using dipole approximation and length gauge, in region 2 we have
\begin{equation}
H_{2}(x,t)=\frac{p^{2}}{2m}+exF(t),
\label{H2}
\end{equation}
where the external electric field is assumed to have oscillating time dependence: $F(t)=F_{0}\cos{(\omega t)}.$ At the boundary of regions 2 and 3 ($x=L$), $F(t)$ becomes zero, and, correspondingly, the gradient of the potential also vanishes in region 3. Since the potential has to be continuous, we can write
\begin{equation}
H_{3}(t)=\frac{p^{2}}{2m}+eLF(t)=H_{2}(L,t).
\label{H3}
\end{equation}
Note that the time dependence of the spatially constant potential corresponds to an overall, time dependent phase factor for any wave function, and we still have free propagation in region 3. In region 4, where the second interaction takes place, the potential of the laser field is superimposed on the oscillation of $H_{3}$:
\begin{equation}
H_{4}(x,t)=H_{3}(t)+e\tilde{x}\tilde{F}(t),
\label{H4}
\end{equation}
where $\tilde{x}=x-d-L,$ and the electric field has the same amplitude and frequency as in region 2: $\tilde{F}(t)=F_{0}\cos {(\omega t+\varphi _{0})}.$ As we shall see later, the phase difference $\varphi _{0}$ (which is zero for the example shown in Fig.~\ref{fig1}) can be used to control the transmission probability. Finally, the Hamiltonian in region 5 describes free propagation again, $H_{5}$ is spatially constant but oscillates in time, with its potential part being equal to that of $H_{4}$ evaluated at the boundary of regions 4 and 5.

\section{Floquet solutions}

Since we consider time-periodic Hamiltonians, it is plausible to use Floquet theory \cite{F883,S65,Kohler_Floq2005}. That is, in all regions, we are seeking solutions of the time dependent Schr\"{o}dinger equation in the form
\begin{equation}
\psi _{j}(x,t)=e^{-\frac{iE_{j}t}{\hbar }}\Phi _{j}(x,t),
\label{floq}
\end{equation}
where $E_{j}$ is the Floquet quasienergy of "channel" $j$, where the index $j$ is an integer. The Floquet state $\Phi_{j}(x,t)$ is a periodic function: $\Phi_{j}(x,t)=\Phi_{j}(x,t+\tau)$ with a time period of $\tau=2\pi /\omega$. Since the global Hamiltonian (that can be obtained by applying $H_{1},\ldots ,H_{5}$ in their respective domains, i.e., regions $1,\ldots ,5$) is periodic in time, global Floquet-type solutions (that are defined on the whole $x$ axis) of the time dependent Schr\"{o}dinger equation exist. We can express $\Phi _{j}(x,t)$ as a Fourier series
\begin{equation}
\Phi_{j}(x,t)=\sum_{n=-\infty}^{\infty}\chi_{n}^{(j)}(x)e^{-in\omega t},
\label{fourier}
\end{equation}
where the expansion functions $\chi_{n}^{(j)}(x)$ do not depend on time.

Particularly, even in region 1, where the potential is zero, there exist solutions of the form of (\ref{floq}). We demonstrate this by taking into account also the corresponding boundary condition for $x<0$. For that region, the physical situation requires to have a superposition of the incident right propagating free plane wave as given by Eq.~(\ref{psiin}) plus the reflected waves. As one can easily see, $E_{0},$ the energy of the incoming wave, should be one of the Floquet quasienergies $E_{j}.$ Therefore, the wave function can be written as
\begin{equation}
\Psi_{1}(x,t)=\psi_{in}(x,t)+\sum\limits_{n}r_{n}e^{-ik_{n}x}e^{-i\omega
_{n}t},
\label{psi1}
\end{equation}
where the wave numbers corresponding to different Floquet quasienergies are defined as follows
\begin{equation}
k_{n}=\sqrt{\frac{2mE_{n}}{\hbar ^{2}}},
\label{kn}
\end{equation}
and $E_{n}=E_{0}+n\hbar \omega $, where $n=(...,-2,-1,0,1,2,...)$. The frequencies appearing in the time evolution are
\begin{equation}
\omega _{n}=E_{n}/\hbar =E_{0}/\hbar +n\omega .
\end{equation}
We note here, that below a certain integer $n$, the wave number $k_{n}$ will be purely imaginary, which describes evanescent waves, with decaying amplitude as $x\rightarrow -\infty $. The Floquet quasienergies (or frequencies) with different integers $n$ serve as a plane wave basis set of the wave functions. In the following, we will construct the solutions of all the other regions using this basis set.

The fundamental solutions of the time-dependent Schr\"{o}dinger equation with the Hamiltonian given by Eq.~(\ref{H2}) are the well-known Gordon-Volkov states \cite{gordon,volkov} in the length gauge:
\begin{align}
\psi_{q,\varphi _{0}}^{V}(x,t)& =e^{-i\left[ \alpha \sin{2(\omega
t+\varphi_{0})}-\beta (q)\cos{(\omega t+\varphi _{0})}\right] }  \notag \\
& \times e^{-i\gamma x\sin{(\omega t+\varphi _{0})}}e^{i\left( qx-\frac{%
\mathcal{E}(q)}{\hbar }t\right) }.
\label{volk}
\end{align}
Here, we have used similar notations as in Ref.~\cite{varro1998}:
\begin{equation}
\mathcal{E}(q)=\frac{\hbar ^{2}q^{2}}{2m}+U_{p},\qquad U_{p}=\frac{%
e^{2}F_{0}^{2}}{4m\omega ^{2}},
\label{pond}
\end{equation}
and
\begin{equation}
\alpha =-\frac{U_{p}}{2\hbar \omega },\qquad \beta (q)=-\frac{eqF_{0}}{%
m\omega ^{2}},\qquad \gamma =\frac{eF_{0}}{\hbar \omega }.
\end{equation}
The ponderomotive potential $U_{p}$ is the classical cycle-averaged energy of the free charged particle in a sinusoidal oscillating electric field. According to (\ref{pond}), the wave number $q$ is related to $\mathcal{E}(q)$ through the dispersion relation
\begin{equation}
q=\sqrt{\frac{2m(\mathcal{E}-U_{p})}{\hbar ^{2}}}.
\label{wn_q}
\end{equation}
and each $\mathcal{E}(q)$ is doubly degenerate due to the two possible propagation directions.

In region 2, the wave function for a given energy reads
\begin{equation}
\psi _{2}(x,t)=a\psi _{q,0}^{V}(x,t)+b\psi _{-q,0}^{V}(x,t).
\end{equation}
More generally, we can write
\begin{equation}
\Psi _{2}(x,t)=\sum_{n}\left[ a_{n}\psi _{q_{n},0}^{V}(x,t)+b_{n}\psi
_{-q_{n},0}^{V}(x,t)\right] ,
\end{equation}%
where $a_{n}$ and $b_{n}$ denote the coefficients of right (or decaying) and
left propagating (or rising) wave modes, respectively. In order to achieve
the Fourier series form (\ref{fourier}), we use the Jacobi-Anger identities
\begin{equation}
\label{jacang1}
e^{i x \sin{\theta}}=\sum\limits_{s=-\infty}^{\infty}J_s(x) e^{i s \theta},
\end{equation}
\begin{equation}
\label{jacang2}
e^{i x \cos{\theta}}=\sum\limits_{s=-\infty}^{\infty} i^s J_s(x) e^{i s \theta},
\end{equation}
where $J_{s}$ denotes the Bessel function of the first kind
\cite{GR}. As a result, the wave function in region 2 reads
\begin{align}
& \Psi _{2}(x,t)=\sum_{n,p,s}J_{s}(\alpha )i^{2s-n+p}\{a_{p}J_{2s-n+p}[\beta
(q_{p})]e^{iq_{p}x}+  \notag \\
& +b_{p}J_{2s-n+p}[\beta (-q_{p})]e^{-iq_{p}x}\}e^{-i\gamma x\sin {(\omega t)%
}}e^{-in\omega t}.
\label{psi2}
\end{align}
We note that the factor $\exp {[-i\gamma x\sin {(\omega t)}]}$ can also be expanded using the Jacobi-Anger formulas leading to one more summation index in Eq.~(\ref{psi2}). However, since in the fitting equations (see the Appendix) this factor is always 1 or canceled out, we omit the expansion for the sake of brevity.

After the first interaction region, the particle is in region 3, outside the influence of the laser-field. It propagates further in this intermediate zone with an altered energy due to the effect of the electric field in region 2. This corresponds to a time-periodic oscillating rectangular potential (see e.g.~\cite{LR99,DiracSCBF}). Since the commutator  $\left[ H_{3}(t),H_{3}(t^{\prime })\right] =0$, the solution of the time-dependent Schr\"{o}dinger equation with the Hamiltonian (\ref{H3}) can be constructed by direct time-domain integration. Considering the double degeneracy of the wave numbers, as well as the desirable Floquet form, the total wave function in region 3 reads
\begin{align}
\Psi _{3}(x,t)& =\sum_{n}(u_{n}e^{ik_{n}x}+v_{n}e^{-ik_{n}x}) \notag \\
& \times e^{-i\gamma L\sin {\omega t}}e^{-in\omega t}.
\label{psi3}
\end{align}

Describing the second interaction of the particle wave with the electric field (in region 4) is very similar to the first one in region 2. We obtain
\begin{align}
\Psi_4(x,t)&=\sum_n \left[c_n \psi_{q_n,\varphi_0}^{V}(x,t)+d_n
\psi_{-q_n,\varphi_0}^{V}(x,t)\right]  \notag \\
&\times e^{-i \gamma L \sin{\omega t}} e^{-i n\omega t},
\end{align}
where $c_n$ and $d_n$ correspond to right and left propagating modes, respectively. Using the Jacobi-Anger identities again, we can transform the wave function into Floquet form
\begin{align}
\Psi_4(x,t)&=\sum_{n,p,s} J_s(\alpha) i^{2s-n+p}\lbrace c_p
J_{2s-n+p}[\beta(q_p)] e^{i q_p x}+  \notag \\
&+d_p J_{2s-n+p}[\beta(-q_p)] e^{-i q_p x}\rbrace e^{-i\gamma x \sin{(\omega
t+\varphi_0)}} \notag \\
&\times e^{-i\gamma L \sin{(\omega t})} e^{-i (n-p)\varphi_0} e^{-i n\omega
t}.
\end{align}

Finally, the wave function in region 5 consists of free modes with two additional oscillating phase factors:
\begin{align}
\Psi_5(x,t)&=\sum_n t_n e^{i k_n x} e^{-i\gamma L \sin{(\omega t+\varphi_0)}} \notag \\
&\times e^{-i\gamma L \sin{(\omega t)}} e^{-i n\omega t}.
\label{psi5}
\end{align}
These are the transmitted waves which are propagating right, see Fig.~\ref{fig1}.

\bigskip

We constructed wave functions with purely exponential time dependences using Floquet theory. For practical reasons, we use local space coordinates, i.e., the origin is redefined in each region [e.g., see the introduction of $\tilde{x}$ in Eq.~(\ref{H4})]. The origins of the first and the second regions coincide, which is the zero of the global coordinate system (see Fig.~\ref{fig1}). Therefore, the continuity boundary conditions for the wave functions and for their derivatives read
\begin{align}
\begin{split}
\Psi_1(0,t)=\Psi_2(0,t),&\quad\Psi_2(L,t)=\Psi_3(0,t), \\
\Psi_3(d,t)=\Psi_4(0,t),&\quad\Psi_4(L,t)=\Psi_5(0,t).
\end{split}
\label{bndCond1}
\end{align}
\begin{align}
\begin{split}
\partial_x\Psi_1(0,t)=\partial_x\Psi_2(0,t),\
\partial_x\Psi_2(L,t)=\partial_x\Psi_3(0,t), \\
\partial_x\Psi_3(d,t)=\partial_x\Psi_4(0,t),\
\partial_x\Psi_4(L,t)=\partial_x\Psi_5(0,t).
\end{split}
\label{bndCond2}
\end{align}
See the Appendix for the actual fitting equations. Considering these boundary conditions for each Floquet channel, we obtain an infinite system of linear equations for the unknown coefficients. However, since the Bessel functions (appearing in the expressions of the wave functions) decrease as a function of their index, it is sufficient to take only a finite number of frequencies into account. In the next section, we present the results based on this model.

\section{Results and discussion}

We investigate time-averaged transmission spectra for different field parameters using the previously introduced Floquet theory. As we shall see, the main features of the transmission probability as a function of the energy of the incoming particles can be understood using an appropriate static model. More details can be seen by exploring the role of the different scattering channels. Beside the time-averaged quantities, we also study the dynamics of the wave-packets generated by the interaction of the particle wave with the optical fields.

\subsection{Cycle-averaged transmission probability}

The usual probability current density in one dimension is defined as
\begin{equation}  \label{current}
j(x,t)=\frac{\hbar}{m}\hbox{Im}\left[\Psi^*(x,t)\frac{\partial\Psi(x,t)}{%
\partial x}\right].
\end{equation}
Time-dependent transmission (reflection) probabilities are given by the ratio of the transmitted (reflected) current to the incoming one. By using Eqs.~(\ref{psiin}) and (\ref{psi5}), one can realize that the time dependence of the probability currents contain factors $\exp[-i (n-m)\omega t]$, i.e., the transmission probability is a periodic function of time.

As we noted before, integer indexes $n$, that correspond to imaginary wave numbers $k_n$, mean evanescent modes. It can be readily seen from Eq.~(\ref{current}), that these waves do not carry probability currents, neither do they make any contributions to the transmission probability. The cycle-averaged current components of reflection and transmission (normalized to the incoming current) are given by
\begin{equation}
j_n^R=\frac{k_n}{k_0}|r_n|^2, \qquad j_n^T=\frac{k_n}{k_0}|t_n|^2,
\end{equation}
where wave numbers $k_n$ are defined in Eq.~(\ref{kn}). Thus, the cycle-averaged reflection and transmission probabilities read
\begin{equation}  \label{RT}
\left\langle R \right\rangle =\sum\limits_{n=n_0}^{\infty}j_n^R, \quad
\left\langle T \right\rangle =\sum\limits_{n=n_0}^{\infty}j_n^T,
\end{equation}
where $n_0$ is the lowest Floquet index, for which the wave number $k_n$ is real.

The total probability must be conserved, which is formulated in one dimension as
\begin{equation}
\frac{\partial\rho(x,t)}{\partial t}+\frac{\partial j(x,t)}{\partial x}=0,
\end{equation}
where we define the probability density as $\rho=\Psi^*\Psi$. Accordingly,
\begin{equation}  \label{R+T}
\left\langle R \right\rangle + \left\langle T \right\rangle = 1
\end{equation}
should always hold for any system parameters \cite{saczuk2003,DiracSCBF}. This condition also serves as an accuracy indicator of our calculations. The infinite system of equations has to be truncated in accordance with an acceptable arbitrary limit of error. For the results to be presented in the following, $|1-\left\langle T \right\rangle-\left\langle R \right\rangle|\leq 10^{-6}$ is chosen. This condition can always be met by increasing the number of modes (Floquet channels) that are taken into account. For the parameters we used, the highest order Floquet index was around $25$.

%%%%%%%%%%%%%%%%%%%%%%%%%%%%%%%%%%%%%%%%%%%%%%%%
\begin{figure}[tbh]
\includegraphics[width=8.0cm]{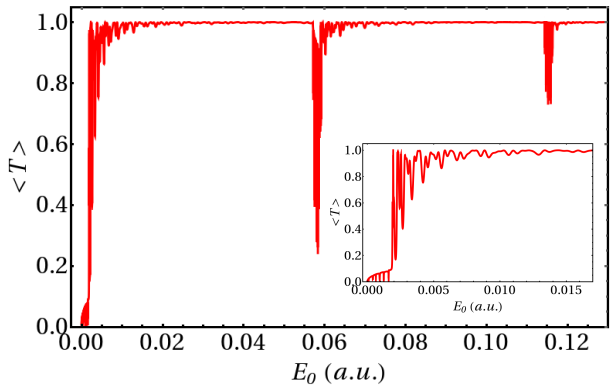}
\caption{Cycle-averaged transmission probability $\left\langle T\right\rangle$ as a function of the energy of the incoming particle $E_0$. Parameters in atomic units are $F_0=0.00488$, $L=200$, $d=400$, $\protect\varphi_0=\protect\pi$ and $\protect%
\omega=0.057322$ corresponding to a wavelength of $800\,nm$.}
\label{CycT_10nm}
\end{figure}
%%%%%%%%%%%%%%%%%%%%%%%%%%%%%%%%%%%%%%%%%%%%%%%%

In Fig.~\ref{CycT_10nm}, the cycle-averaged transmission probability $\left\langle T \right\rangle$ is plotted as a function of $E_0$, which denotes the energy of the incoming particle. The parameters correspond to localized optical fields ($L\approx10\,nm$, $d\approx20\,nm$) that can be realized experimentally \cite{kruger2011}. As we can see, in this case the transmission spectrum is complex, there are numerous maxima and minima. The most important parameter here is the ratio of the de Broglie wavelength $\lambda_{dB}=2\pi/k_0$ of the incoming particle and the separation of the interaction regions, $d.$ When, e.g.,  $d=n (\lambda_{dB}/2),$ with a large integer $n,$ increasing $E_0$ by a few percent of its initial value can result in a similar ratio of $d$ and $\lambda_{dB},$ with $n$ replaced by $n+1.$ Since the length of $d$ corresponds again to an integer multiple of $\lambda_{dB}/2,$ the interference pattern will be approximately the same. Therefore, for $d\gg \lambda_{dB},$ whenever we see e.g., a peak in the transmission spectrum, it will be repeated multiple times within a short energy interval. This is the case for the parameter range shown in  Fig.~\ref{CycT_10nm}. In order to simplify the interpretation, in the following we consider the regime where $d$ is not too large in comparison to $\lambda_{dB},$ which leads to a less complex interference pattern, the understanding of which can be straightforwardly transferred to different parameter regimes as well. Let us note that by increasing $d,$ it is not only the number of the peaks in the transmission spectra that is seen to increase, but the widths of the individual peaks also change. Larger separation of the interaction regions results in narrower peaks, which is general for Ramsey-like setups, and allows e.g., precise energy filtering. Besides these quantitative differences, according to our results, the physical picture that explains the interference pattern for $d\geq\lambda_{dB}$ is still valid for $d\gg \lambda_{dB},$ and all the results, including the $\varphi_0$ dependence of the transmission, hold also in this case.

%%%%%%%%%%%%%%%%%%%%%%%%%%%%%%%%%%%%%%%%%%%%%%%%
\begin{figure}[tbh]
\includegraphics[width=8.0cm]{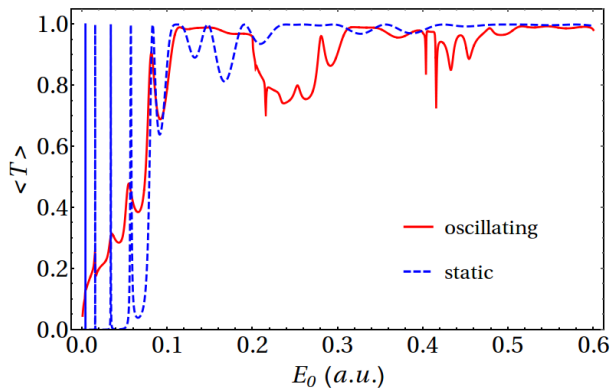}
\caption{Cycle-averaged transmission probability $\left\langle T \right\rangle$ (red solid line) and transmission probability for the corresponding static double barrier (blue dashed line, see the main text for more details) as a function of the energy of the incoming particle $E_0$ in atomic units. Parameters are $F_0=0.1$, $L=10$, $d=30$, $\protect%
\omega=0.2$ and $\protect\varphi_0=\protect\pi$. For low energies, the cycle-averaged transmission probability has peaks at the same input energies where transmission resonances occur in the static double barrier model.}
\label{figTcycle}
\end{figure}
%%%%%%%%%%%%%%%%%%%%%%%%%%%%%%%%%%%%%%%%%%%%%%%%

Fig.~\ref{figTcycle} shows the cycle-averaged transmission probability $\left\langle T \right\rangle$ as a function of the energy of the incoming particle. (This is denoted by red solid line, while the meaning of the dashed blue line will be explained in the next subsection.) Increasing this energy means that the transmission probability approaches unity as expected. However, before the saturation happens, a system of transmission peaks and dips are observed at particular values of $E_0$. The details of the transmission spectrum will be explained in successive steps in the next subsections.

\subsection{Scattering resonances}

As we can see, in regions 2 and 4 of Fig.~\ref{fig1} (where the oscillating field is localized) the wave numbers defined by Eq.~(\ref{wn_q}) are exactly the same as in the case of a static rectangular potential barrier with a height of the ponderomotive potential $U_p$ defined in Eq.~(\ref{pond}), where $U_p\geq0$ holds for any charged particle. Along this line, as a first approximation, we can replace the two oscillating linear potentials with a static, symmetric rectangular double-barrier system \cite{Yamamoto1987,DK2010}. The first consequence of this model is a correct prediction for the overall $E_0$ dependence of the time averaged transmission probability: when $E_0\ll U_p,$ $\langle T\rangle$ is close to zero, while for input energies considerably above $U_p,$ it is not far from unity. (See the dashed blue line in Fig.~\ref{figTcycle}.) Clearly, the transition between $\langle T\rangle=0$ and $1$ takes place around $U_p.$

Additional aspects of the transmission spectra can also be understood using the static model described above. By determining the transmission probability of an incoming particle of energy $E_0$ for two rectangular barriers of height $U_p$ as a static scattering process, we obtain multiple sharp resonances at certain energies, as shown in Fig.~\ref{figTcycle} by the blue dashed line.

Transmission resonances are generally related to the presence of bound states, quasi-bound states or other localized solutions. In order to understand the transmission spectra shown in Fig.~\ref{figTcycle}, it is instructive to find the eigenstates and eigenenergies of the static double barrier system. To this end, we consider a discretized version of the model, where the time-independent Schr\"{o}dinger equation is solved with two spatially separated potential barriers of height $U_p$. Periodic boundary condition is used: the Hamiltonian matrix is constructed in such a way that the rightmost grid point is connected to the leftmost one.

%%%%%%%%%%%%%%%%%%%%%%%%%%%%%%%%%%%%%%%%%%%%%%%%
\begin{figure}[tbh]
\includegraphics[width=8.0cm]{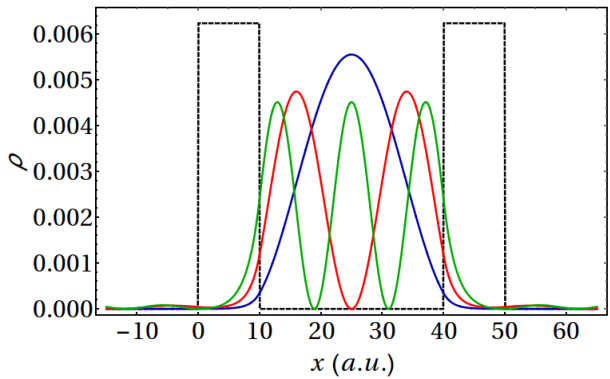}
\caption{Probability densities of the first three localized states in the double barrier system. Localized states with increasing eigenenergies are
denoted by blue, red, and green lines, respectively. As a reference, the
potential barriers are drawn by black dashed lines.}
\label{locstates}
\end{figure}
%%%%%%%%%%%%%%%%%%%%%%%%%%%%%%%%%%%%%%%%%%%%%%%%

Fig.~\ref{locstates} shows the probability densities calculated for three such eigenstates which are found to be localized between the two potential barriers that are indicated by black dashed lines in the figure. These states correspond to the same energies, where the transmission spectrum (in Fig.~\ref{figTcycle}) has pronounced peaks. In other words, the reason for the transmission resonances observed for the static potential barriers is the existence of these localized states.

Returning to Fig.~\ref{figTcycle}, now it is clearly seen, that the previously introduced scattering problem with oscillating potential also has transmission resonances around these particular energies. That is, the static model can be viewed as a first approximation for low energies (below $E_0=\hbar\omega$). However, as it is clear by comparing the red and blue curves in Fig.~\ref{figTcycle}, for higher energies, the oscillation of the potential results in a structured transmission spectrum that cannot be explained by the static model. The physical processes determining this part of the spectra will be examined in the following subsection.

\subsection{Shifted, "multiphoton" resonances}

%%%%%%%%%%%%%%%%%%%%%%%%%%%%%%%%%%%%%%%%%%%%%
\begin{figure}[tbh]
\includegraphics[width=8.0cm]{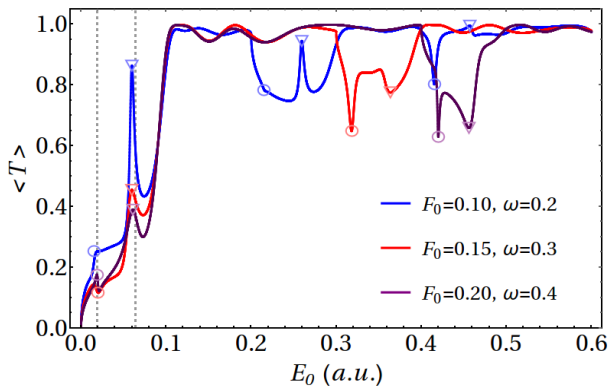}
\caption{Cycle-averaged transmission probabilities $\left\langle T \right\rangle$ as a function of the incoming particle's energy $E_0$ in atomic units. Parameters: $L=10$, $d=10$, and $\protect\varphi_0=\protect\pi$. The pale circles and triangles correspond to the first and second scattering resonances, respectively.}
\label{figTcycle_fixUp}
\end{figure}
%%%%%%%%%%%%%%%%%%%%%%%%%%%%%%%%%

As a next step, we investigate the scattering process by varying the parameters so that the pondoromotive energy $U_p$ defined by (\ref{pond}) is kept constant. In this way the static model introduced in the previous subsection is unchanged (by definition), and we can explore effects beyond this approximation. (In other words, we are to explain the difference between the two curves in Fig.~\ref{figTcycle}.) Fig.~\ref{figTcycle_fixUp} shows that for different electric field amplitudes and angular frequencies of oscillation, the locations of the resonances (below $
E_0=\hbar\omega$) stay approximately the same. In more details, the parameters in this figure correspond to only two static scattering resonance energies (at $E_0=0.01926$ and $0.0643$ a.u.), which are denoted by gray dashed lines as references. As we can see, the static barriers indeed mean good approximation for the expected
resonance energies when $E_0<\hbar\omega$.

For larger input energies, however, there are even more peaks and dips in the transmission probability for the oscillating case. In Fig.~\ref
{figTcycle_fixUp}, circles and triangles correspond to the static scattering resonance energies shifted by $n\hbar\omega.$ These resonances can be explained by noticing, that after losing an integer multiple of the energy quanta $\hbar\omega$, the energy of the scattered particle coincides with the energy of one of the previously shown localized states. In this sense, the pale circles and triangles correspond to the first and second scattering resonance energies, respectively. Note that this explains the energy value at which these resonances appear, but the behavior of the transmission probabilities at these energies (e.g., whether we experience a peak or a
dip) needs a more detailed description, see the next subsection.

\begin{figure}[tbh]
\includegraphics[width=8.0cm]{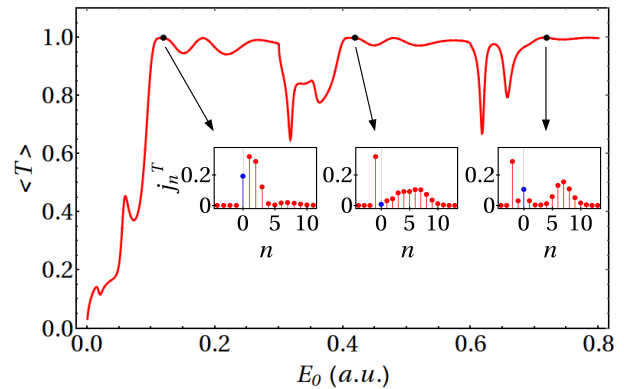}
\caption{Cycle-averaged transmission probability $\left\langle T
\right\rangle$ as a function of incoming particle's energy $E_0$ in atomic
units. Parameters: $F_0=0.1$, $L=10$, $d=10$, $\protect\omega=0.3$ and $%
\protect\varphi_0=\protect\pi$. All the insets show the individual current
contributions $j_n^T$ of the Floquet channels to the transmission probability for different values of $E_0.$}
\label{TE0+inset}
\end{figure}

As an additional interesting feature, Fig.~\ref{TE0+inset} shows the transmission probability as well as the contributions of every single Floquet channel to it at different energies. Although the transmission probability is almost unity in the three specified energies, they correspond to entirely different probability current distributions as it is shown by the insets of Fig.~\ref{TE0+inset}. The blue dot in the insets marks the probability current of the central Floquet channel ($n=0$).

When the incoming energy $E_0$ reaches $\hbar\omega$, the scattering channel $n=-1$ opens and can also contribute to the transmission probability. The
same happens after every single additional energy quantum $\hbar\omega$: a previously evanescent wave mode transforms into a propagating one. This phenomenon is due to the emission of "photons", where the particle can transmit energy to its environment. Therefore, the probability currents $
j_n^T$ can be also called the "multiphoton" components of transmission.

\subsection{Phase dependence of the transmission}

So far, Figs.~\ref{figTcycle}, \ref{figTcycle_fixUp} and \ref{TE0+inset} show transmission spectra when the two localized electric fields have a phase difference $\varphi_0$ of $\pi$, which corresponds to a symmetric oscillating trapezoid potential. In the following, we inspect the phase difference dependence of the scattering process.

\begin{figure}[tbh]
\includegraphics[width=8.0cm]{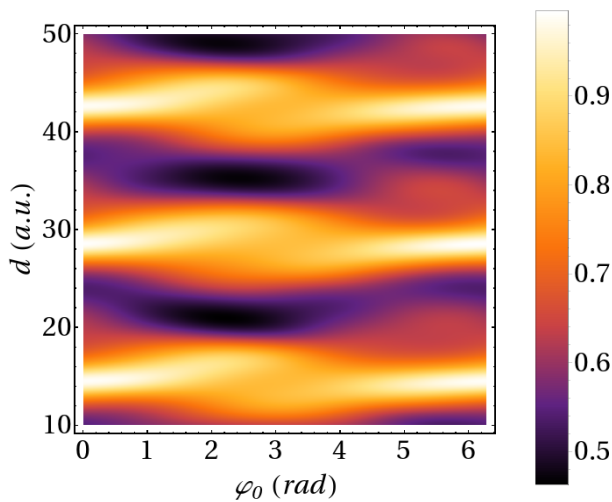}
\caption{Transmission probabilities as a function of the phase difference $\protect\varphi_0$ for different lengths $d$ of region 3. System parameters: $%
E_0=0.025$, $\lambda_{dB}=28.0993$, $F_0=0.1$, $L=10$, $\protect\omega=0.2$.}
\label{Tfi0}
\end{figure}

Fig.~\ref{Tfi0} shows the transmission probability as a function of $\varphi_0$ for different separation distances (denoted by $d$) of the optical fields. The quasiperiodicity of the time-averaged transmission probability as a function of $d$ (with a period of $\lambda_{dB}/2$) clearly shows the fact that we have already mentioned earlier: for two values of $d$ for which $2 d_1/\lambda_{dB}$ and $2 d_2/\lambda_{dB}$ differs by only an integer, the interference pattern is very similar, leading to similar transmission probabilities. According to our calculation, when all other parameters are fixed, $\langle T \rangle$ can change $50$\% as a function of  $\varphi_0$, and this behavior is observable also for experimentally relevant parameter ranges.

In order to understand the detailed role of the phase difference in the scattering process, we analyze the space and time dependence of the probability current density. Generally, due to the population of various Floquet channels, the solution will obviously not be monoenergetic, propagating wavepackets emerge. As a physical picture, we may consider that the wave-packets generated in region 2 approach the second optical field, where, depending on the
relative phase difference $\varphi_0$, the slope of the potential experienced by the wave-packets will be different. In other words, the wave packets entering region $4$ will either experience an "attractive" potential that forces them to move towards region $5$ (and consequently increase the transmission probability), or a "repulsive" one leading to reflection.  As a consequence, focusing on transmission resonances, we can observe that transmission peaks can transform into dips and vice versa as we sweep through $\varphi_0$. Clearly, this is only a first approach (since, e.g., oppositely moving wave-packets in region $3$ can interfere), but it can capture the essential mechanism beyond the $\varphi_0$ dependence of $\langle T \rangle.$

\begin{figure}[tbh]
\includegraphics[width=8.0cm]{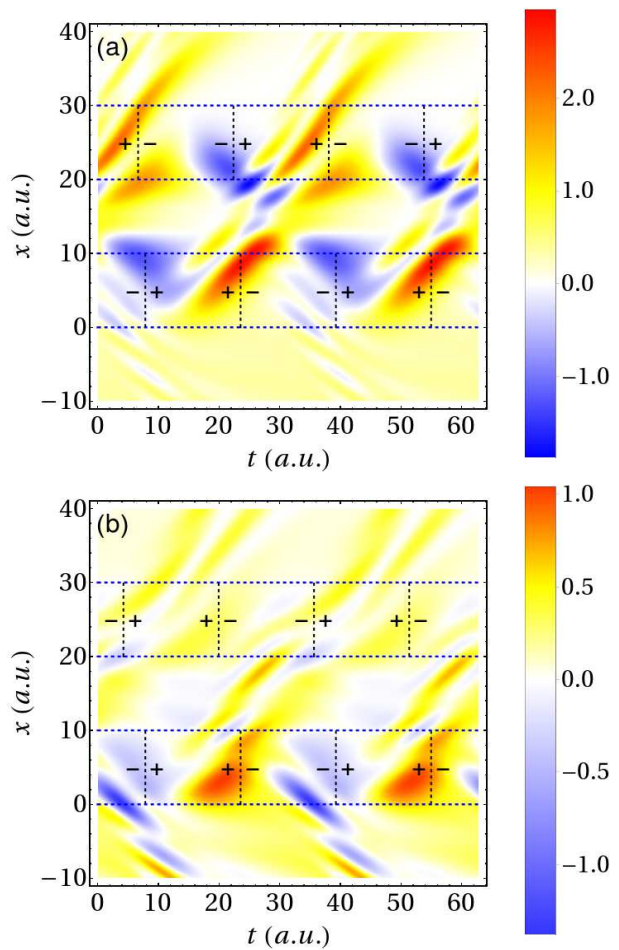}
\caption{Density plot of $j(x,t)$ as a function of time and coordinate $x.$ For parameters $E_0=0.06$, $F_0=0.1$, $L=10$, $d=10$, and $\protect\omega=0.2,$ panels (a) and (b) correspond to the maximal and minimal transmission probabilities, respectively. Numerically: $\langle T \rangle = 0.8941$ at $\protect\varphi_0=\protect 3.3772$ for panel (a), and $\langle T \rangle = 0.392$ at $\protect\varphi_0=\protect 0.7226$ in panel (b). Horizontal dashed lines indicate the boundaries of the different regions, vertical ones correspond to the zeros of the classical force exerted on the charged particle. The sign of this force is also shown in the various space-time regions.}
\label{5regcurr}
\end{figure}

As an illustration, Fig.~\ref{5regcurr} (a) and (b) show density plots of $j(x,t)$ for two different $\varphi_0$ values (all other parameters are the same, see the caption). Subfigures (a) and (b) correspond to the maximum and minimum of $\langle T \rangle,$ respectively. In both cases, the ripples in region 1 are related to the interference of the incoming and reflected waves. Two optical cycles are shown, and the $\pm$ signs in the interaction regions show the sign of the classical force that corresponds to the oscillating potential.  As we can see, when the transmission probability is minimal, the most pronounced wave packet reaches region $4$ in a time interval when the oscillating potential repels it. On the other hand, as it is shown by Fig.~\ref{5regcurr} (a), the maximum of $\langle T \rangle$ corresponds to the case when the wave packet in the second interaction region is pushed towards region $5.$ Note that for higher input energies and stronger optical fields more structured wave-packets are generated in the outermost region. Additionally, when $E_0\gg U_p$ (or $E_0\ll U_p$) $\langle T \rangle$ is very close to unity (or zero) and consequently cannot have strong $\varphi_0$ dependence. Therefore, in order to control the transmission by changing $\varphi_0,$ the parameters of the electric fields must correspond to a ponderomotive potential close to the characteristic kinetic energy, $E_0,$ of the particle beam.

As a possible application, let us emphasize that the time-averaged transmission probability can strongly depend on the phase difference $\varphi_0$ also at the transmission resonances. For large enough separation of the interaction regions, these resonance peaks are narrow, and consequently, for a realistic, non-monoenergetic particle beam, they can serve as narrow band energy filters. More interestingly, the properties of these energy filters can be controlled by changing only $\varphi_0,$ without modifying any other parameters of the experimental setup.

\section{Summary and conclusions}

We presented a non-relativistic time-periodic scattering problem where a
charged particle, e.g., an electron was assumed to be scattered on two
spatially localized time-periodic optical fields. Considering dipole
approximation and using Floquet theory, the cycle-averaged transmission
probabilities were calculated with different system parameters. Results
showed a very sophisticated spectrum, which was explained in successive
steps. First, we recognized that a double potential barrier system (with
barrier heights being equal to the ponderomotive potentials) serves as a
fair approximation for low energies. We determined the energies of the
scattering resonances in the static model, and identified them in the
spectrum of the time-dependent model. We also explained the additional
resonances occurring at higher energies through the behavior of the
probability currents belonging to the Floquet channels. Finally, we
explained the phase difference dependence of the transmission probability by
inspecting the temporal behavior of the generated wave-packets.

The results presented here point out how optical fields can control moving charged particles. Specifically, we determined the parameter range in which the mere phase difference of the
optical fields can control the transmission probability. Although we used the context of a beam propagating in free space, understanding the basic phenomena that govern interferometric processes induced by separated fields is crucial also from the viewpoint of ultrafast, light-induced electronics, i.e., when the charged particles move in a solid. Although our model focuses on the most important, qualitative aspects of the interaction, it can provide an adequate first approach to more complex systems as well. With acceptable increase of numerical costs, our method can also treat two dimensional problems or bichromatic excitation. As an important generalization, the spatial dependence of the exciting fields can also be taken into account.

\section*{Acknowledgments}

Partial support by the ELI-ALPS project is acknowledged. The ELI-ALPS project (GINOP-2.3.6-15-2015-00001) is supported by the European Union and co-financed by the European Regional Development Fund. The work was also supported by the European Social Fund and the 'Sz\'{e}chenyi 2020' program under contract EFOP-3.6.2-16-2017-00005.

\section*{Appendix}

As an example, the equation describing the continuity of the wave function at the boundary of the first and second region, for the $n$th Floquet channel reads
\begin{align}
\delta_{n0}+r_n=&\sum_{p,s} J_s(\alpha) i^{2s-n+p}\lbrace a_p
J_{2s-n+p}[\beta(q_p)]+  \notag \\
&+b_p J_{2s-n+p}[\beta(-q_p)]\rbrace,
\end{align}
where the factor $\exp [-i\gamma x \sin{(\omega t)}]$ equals unity at $x=0.$
The fitting equation originating from the continuity of the derivatives is a bit more complex. The wave function $\Psi_2(x,t)$ has two coordinate-dependent factors. The derivatives read as follows:
\begin{align}
&\frac{\partial}{\partial x}\left\{ e^{-i\gamma x \sin{(\omega t)}} e^{\pm i
q_p x}\right\}=e^{-i\gamma x\sin{(\omega t)}}  \notag \\
& \times e^{\pm i q_p x}\left[-i\gamma\sin{(\omega t)}\pm i q_p\right].
\end{align}
At the boundary $x=0$, the derivative is
\begin{equation}
-\frac{\gamma}{2}\left(e^{i\omega t}-e^{-i\omega t}\right)\pm i q_p.
\end{equation}
Therefore, after shifting the summation indices, the fitting equations for the derivatives read
\begin{align}
&i k_0 \delta_{n0}+r_n (-i k_n)=\sum_{p,s} J_s(\alpha) i^{2s-n+p+1}\times
\notag \\
&\times\left\{ a_p J_{2s-n+p}[\beta(q_p)]\left(q_p+\frac{\gamma (2s-n+p)}{%
\beta(q_p)}\right)\right.+  \notag \\
&\left.+b_p J_{2s-n+p}[\beta(-q_p)]\left(-q_p+\frac{\gamma (2s-n+p)}{%
\beta(-q_p)}\right)\right\}.
\end{align}

When fitting at the boundary of regions 2 and 3, $\exp{(-i\gamma
L\sin{\omega t})}$ cancels out. This factor and $\exp{[-i\gamma L\sin{(\omega t+\varphi_0)}]}$ in region 4 and 5 are also found to be trivial in the
corresponding equations.

\end{document}